# Abasy Atlas: A comprehensive inventory of systems, global network properties and systems-level elements across bacteria

Miguel A. Ibarra-Arellano[1,2,§,†], Adrián I. Campos-González[1,2,†], Luis G. Treviño-Quintanilla[3], Andreas Tauch[4], and Julio A. Freyre-González[1,*]

[1]Group of Regulatory Systems Biology, Evolutionary Genomics Program and [2]Undergraduate Program in Genomic Sciences, Center for Genomics Sciences, Universidad Nacional Autónoma de México. Av. Universidad s/n, Col. Chamilpa, 62210. Cuernavaca, Morelos, México

[3]Departamento de Tecnología Ambiental, Universidad Politécnica del Estado de Morelos. Blvd. Cuauhnáhuac 566, Col. Lomas del Texcal, 62550. Jiutepec, Morelos, México

[4]Centrum für Biotechnologie (CeBiTec). Universität Bielefeld, Universitätsstraße 27, 33615. Bielefeld, Germany

**\*Corresponding author:** jfreyre@ccg.unam.mx (JAF-G)

**§Present address:** Department of Molecular Genetics and Microbiology, College of Medicine, University of Florida, 2033 Mowry Road, Gainesville, Florida 32610-0266

**†**These authors contributed equally to this work

**Abstract**

The availability of databases electronically encoding curated regulatory networks and of high-throughput technologies and methods to discover regulatory interactions provides an invaluable source of data to understand the principles underpinning the organization and evolution of these networks responsible for cellular regulation. Nevertheless, data on these sources never goes beyond the regulon level despite the fact that regulatory networks are complex hierarchical-modular structures still challenging our understanding. This brings the necessity for an inventory of systems across a large range of organisms, a key step to rendering feasible comparative systems biology approaches. In this work, we take the first step towards a global understanding of the regulatory networks organization by making a cartography of the functional architectures of diverse bacteria. Abasy (**A**cross-**ba**cteria **sy**stems) Atlas provides a comprehensive inventory of annotated functional systems, global network properties, and systems-level elements (global regulators, modular genes shaping functional systems, basal machinery genes, and intermodular genes) predicted by the natural decomposition approach for reconstructed and meta-curated regulatory networks across a large range of bacteria, including pathogenically and biotechnologically relevant organisms. The meta-curation of regulatory datasets provides the most complete and reliable set of regulatory interactions currently available, which can even be projected into subsets by considering the force or weight of evidence supporting them or the systems that they belong to. Besides, Abasy Atlas provides data enabling large-scale comparative systems biology studies aimed at understanding the common principles and particular lifestyle adaptions of systems across bacteria. Abasy Atlas contains systems and system-level elements for 50 regulatory networks comprising 78,649 regulatory interactions covering 42 bacteria in nine taxa, containing 3,708 regulons and 1,776 systems. All this brings together a large corpus of data that will surely inspire studies to generate hypothesis regarding the principles governing the evolution and organization of systems and the functional architectures controlling them.

**Database URL:** http://abasy.ccg.unam.mx

**Background**

Bacterial regulatory networks (RNs) are responsible for sense stimuli and environmental cues and respond accordingly. In complex environments, they "take composite decisions" to prioritize, for example, the transport and catabolism of carbon sources according to the metabolic preferences of each organism. To accomplish this, RNs composed by thousands of regulatory interactions, must follow well-defined organization principles governing their dynamics. In the last decades of the 20th century, the first levels of gene organization were unveiled as the operon and the regulon.

Currently, a few databases (RegulonDB (1), SubtiWiki (2), DBTBS (3), CoryneRegNet (4), and RegTransBase (5)) extract, by manual curation of literature, the molecular knowledge of gene regulation in different organisms, thus providing an invaluable source of data. Nevertheless, data within these databases never goes beyond the regulon level, whereas cumulated evidence has shown that RNs are complex hierarchical-modular networks (6-11) whose organizational and evolutionary principles are pivotal for determining the dynamics of the cell and still challenging our understanding.

The promising field of synthetic biology aims to apply engineering principles for designing and constructing biological systems and devices. To fulfill this aim, it is crucial to understand the set of organizational principles underpinning how cellular systems interconnect, work and evolve. In synthetic biology, and even in biotechnology, a deep understanding of the alternative regulatory patterns, which have evolved as adaptations to different environments, will inspire better synthetic designs, thus enhancing our ability to optimize a genetic regulatory circuit. To accomplish this, we need an inventory of systems and their properties across a large range of organisms, a key step to rendering feasible comparative systems biology approaches.

From a genomic perspective, identifying clusters of genes working together in a system to achieve a particular physiological function is important. It enables the use of a guilt-by-association strategy to propose hypotheses regarding the physiological function of genes whose annotation by homology analysis is not enough mainly because no homologs have been characterized, e.g. the annotation is "hypothetical predicted/conserved protein".

On the other hand, a recent study has shown that mathematical models, like the ones used in synthetic biology, are highly sensitive to structural variations in the system under study, and that failure to consider the relevance of this structural uncertainty gives rise to bias in the analysis and potentially misleading conclusions (12). Therefore, the ability to infer and delimit the members, inputs and outputs of a system, is key to construct reliable and unbiased models.

In this work, we take the first step towards a global understanding of the RNs organization by making a cartography of the functional architectures of diverse bacteria. Abasy (**A**cross-**ba**cteria **sy**stems) Atlas provides a comprehensive atlas of annotated functional systems (hereinafter also referred as modules), global network properties, and systems-level elements predicted by the natural decomposition approach (NDA) (9,10) for reconstructed and meta-curated RNs across a large range of bacteria, including pathogenically and biotechnologically relevant organisms. This biologically motivated mathematical approach use the global structural properties of a given RN to derive its architecture and classify its genes into categories of systems-level elements (global regulators, modular genes, basal machinery genes, and intermodular genes). The decomposition of diverse RNs into their systems and systems-level elements beyond the regulon allows unraveling the complexity of these networks and provides new insights into the organizational principles governing them. This atlas also provides data enabling large-scale comparative systems biology studies aimed at understanding the common principles and particular lifestyle adaptions of systems across bacteria. The meta-curation of regulatory datasets provides the most complete and reliable set of regulatory interactions currently available, which can even be projected into subsets by considering the force or weight of evidence supporting them or the systems that they belong to.

**The natural decomposition approach: A primer**

There are different levels of description in models of genetic networks (13). The NDA is a large-scale modelling approach characterizing the circuit wiring and its global architecture. It defines a mathematical-biological framework that provides criteria to identify the four classes of systems-level elements shaping RNs: global regulators, modular genes shaping functional systems, basal machinery genes, and intermodular genes. Studies have shown that regulatory networks are highly plastic (14). Despite this plasticity, by applying the NDA our group has found that there are organizational principles conserved by convergent evolution in the RNs of phylogenetically distant bacteria (10). The high predictive power of the NDA has been proven in previous studies by applying it to the phylogenetically distant *Escherichia coli* K-12 (9) and *Bacillus subtilis* 168 (10) and by comparing it with other methods to identify modules (11).

The NDA defines objective criteria (e.g. the $\kappa$-value to identify global regulators) to identify functional systems and systems-level elements in a RN, and rules to reveal its functional architecture by controlled decomposition (Figure 1). It is based on two biological premises (10): 1) A module is a set of genes cooperating to carry out a particular physiological function, thus conferring different phenotypic traits to the cell. 2) Given the pleiotropic effect of global regulators, they must not belong to modules but rather coordinate them in response to general-interest environmental cues.

According to the NDA, every gene in a RN is predicted to belong to one out of four possible classes of systems-level elements, which interrelate in a non-pyramidal, three-layer, hierarchy shaping the functional architecture (9,10) as follows (Figure 2): 1) Global regulators are responsible for coordinating both the 2) basal cell machinery, composed of strictly globally regulated genes, and 3) locally autonomous modules (shaped by modular genes), whereas 4) intermodular genes integrate, at promoter level, physiologically disparate module responses eliciting a combinatorial processing of environmental cues.

**Construction and content**

Abasy Atlas has a three-tier, web-based, client-server architecture. PHP programs are responsible for querying the MySQL database, executing Python scripts, and rendering HTML with CSS webpages containing the extracted information. Interactive network panels display gene neighborhood (i.e., the set of all the genes/complexes regulating or being regulated by a given gene, including the latter) and modules, providing alternative layouts and different visualization options. These network panels are rendered by software running on the web browser of the client using serialized data and were developed using JavaScript, Python, and the state-of-the-art Cytoscape.js library (15).

**Data extraction to create a compendium of regulatory network models across bacteria**

To create a comprehensive compendium of bacterial RNs we first compiled regulatory interaction data for the best-curated RNs available in organism-specific databases: *E. coli* K-12 MG1655 from RegulonDB (1). *B. subtilis* 168 from DBTBS (3) and SubtiWiki (2), and *Corynebacterium glutamicum* ATCC 13032 from CoryneRegNet (4). We then identified six additional regulatory datasets by carrying out a literature review: 1) A recent experimentally supported predicted RN for *B. subtilis* 168 (16). 2) A ChIP-seq based reconstruction of the RN of *Mycobacterium tuberculosis* H37Rv (17) plus three studies also reporting experimentally supported and literature-curated RNs for *M. tuberculosis* H37Rv (18-20). 3) A study reporting a literature-curated RN for *Pseudomonas aeruginosa* PAO1 (21). Finally, we expanded this compendium by retrieving the regulatory dataset for 477 bacteria available in RegTransBase, a database of regulatory interactions based on literature (5).

Regulatory interaction data was obtained in a diversity of formats: tab-delimited flatfiles, Excel spreadsheet (XLS), XML, and HTML. Ad-hoc developed Python scripts parsed and converted regulatory interaction data into a unique intermediate tab-delimited format that we defined. RNs were modeled as directed graphs where nodes represent genes (or regulatory complexes, as we will discuss below) and directed edges stand for regulatory interactions. Regulatory data coming from different sources were reconciled through the process of gene symbols disambiguation and meta-curation as described below.

**Sigma factors**

Sigma factors play an active role in response to stimuli by differentially redirecting the transcriptional machinery towards specific sets of promoters. The competition for a limited resource, the RNA polymerase core, requires a strict regulation on the availability of alternative sigma factors, and even housekeeping sigma factors, to control the delicate interplay among them properly (22). Therefore, sigma factors transcriptionally control and are controlled by the products of other genes, and as such, they are important players in the systems-level organization of RNs. To consider this alternative regulatory layer, we included sigma factors as activators in our RN models.

**Regulatory complexes**

Proteins encoded by different genes can form heteromeric complexes that exhibit regulatory activity. As available regulatory datasets are gene-based, genes encoding subunits of a regulatory complex may regulate the same set of genes, thus duplicating interactions in the RN model. To remove this redundancy, we first curated and compiled a list of heteromeric regulatory complexes for *E. coli* K-12 and *B. subtilis* 168, no data was found for other organisms. Next, for each set of subunits composing a complex, if the subunit-encoding genes are regulated by the same set of regulators and do not exhibit regulatory activity independently, we collapsed the genes encoding subunits into a single node representing the regulatory complex. Otherwise, we added regulatory complexes as nodes that are activated by their subunit-encoding genes. We kept outgoing regulations mediated by a subunit that exhibit regulatory activity independently, and we removed the remaining outgoing regulations from the subunit-encoding genes and added them to the complex according manual curation and removing duplicates.

**Regulation mediated by small RNAs**

Small RNAs have emerged as a novel and relevant regulatory element (23). Given their importance, we retrieved the sRNA-mediated regulatory interactions curated by RegulonDB team and created an additional expanded version of the *E. coli* K-12 RN model based on the all evidences version, the latter because all the evidences supporting sRNA-mediated interactions are weak. Other databases or datasets (e.g. DBTBS and RegTransBase) already include sRNA-mediated interactions, and even attenuators or other RNA-mediated regulatory interactions, and these were added to the atlas.

**Classification of regulatory interactions according to their supporting evidences**

Curation of RNs proceeds by identifying evidences that a certain regulator is able to affect the expression of a gene. These evidences have varying degrees of confidence: some evidences (e.g. footprinting) are strong enough to show that the regulator is able to bind to the upstream region of the regulated gene possibly altering its expression, whereas weak evidences just suggest a hypothetical DNA binding site (e.g. bioinformatics predictions) or a possible indirect effect (e.g. gene expression experiments). Weakly supported interactions that eventually turn out to be spurious act like noise affecting the conclusions drawn from RN analyses and mathematical models disregarding this confounding factor (12). We therefore created two versions of the *E. coli* K-12 and *B. subtilis* 168 (DBTBS) RN models: all evidences (containing both weakly and strongly supported interactions) and strong evidences (accounting for strongly supported evidences only).

RegulonDB team has proposed a classification scheme for the strength of single evidences into 'weak' and 'strong' (24). They define weak evidences as a "single evidence with more ambiguous conclusions, where alternative explanations, indirect effects, or potential false positives are prevalent, as well as computational predictions; for instance, gel mobility shift assays with cell extracts or gene expression analysis". On the other hand, strong evidence is defined as a "single evidence with direct physical interaction or solid genetic evidence with a low probability for alternative explanations; for instance, footprinting with purified protein or site mutation".

Using this classification scheme, we retrieved all evidences supporting each regulatory interaction for the *E. coli* K-12 RN model and computed its confidence level as weak or strong according the following rule: if a regulatory interaction is supported by at least one strong evidence it is classified as strong, otherwise it is weakly supported. For *B. subtilis* 168, we developed an analogous classification scheme by manual curation of the set of experiments supporting regulatory interactions in DBTBS, and we next computed the confidence level for each regulatory interaction as previously described.

**An unambiguous and integrative compendium of gene names, locus tags and synonyms to cope with disagreement among regulatory datasets**

Disagreement in gene names is an important confounding factor in genomic analyses. For instance, when RNs are compared, or any other genomic data is mapped onto them, disagreement in gene names is responsible for missing or duplicated interactions or genes potentially causing bias in conclusions (25). Besides, the herein compiled regulatory datasets use different formats for naming genes: some only use gene names, others only locus tags, and some others a mixture of gene names (including synonyms and database IDs) and locus tags.

To cope with the problem of disagreement in gene names, we compiled a dictionary of gene names, locus tags and synonyms for each gene in the regulatory datasets. This compilation of identity information was carried out by integrating information from UniProtKB (26) and NCBI GenBank (27) for all organisms, and additionally RegulonDB for *E. coli* K-12, SubtiWiki for *B. subtilis* 168 and RegTransBase for organisms obtained from this source (Supplementary File 1). In this step, we also extracted accession IDs to provide cross-links to external databases.

Using this identity information, we developed an algorithm to map gene symbols onto unambiguous canonical-form gene names and to complete the identity information by finding missing locus tags and synonyms. We defined a canonical-form or canonical gene name as a UniProt-proposed gene name (http://www.uniprot.org/help/gene_name) for protein coding genes, or if gene codes for an RNA, an NCBI GenBank gene name, or for the cases of *E. coli* K-12 and *B. subtilis* 168, main names proposed by the corresponding curated database RegulonDB or SubtiWiki, respectively, or a locus tag if no names were available.

We represented locus tags and their corresponding gene names (including synonyms) as the two disjoint sets in a bipartite graph where an edge represents an equivalence relation between sets (Figure 3). We then found that gene names exhibit degeneracy, i.e. some gene names map to different locus tags. Therefore, to remove this ambiguity, we first mapped all gene names onto locus tags by using canonical gene names as keys. Next, we transformed the remaining non-mapped gene names into locus tags by a constrained mapping using the previously constructed bipartite graph as follows (Figure 3): if a gene name maps onto more than one locus tag, the gene name is marked as taboo (forbidden) so that the algorithm does not consider it for future mappings and we conserved it as is for historical consistency, otherwise we mapped it onto that unique locus tag. Finally, we mapped all locus tags onto canonical names, when possible. As a result, we collapsed nodes representing genes that are synonyms into a single node and removed duplicated edges in RN models.

**Meta-curation of regulatory network models**

The development of high-throughput interaction discovery methodologies (e.g. ChIP-seq) and gene-expression-based methods for inference of RNs has produced novel regulatory datasets with high genomic coverage. Examples of this are the datasets for *B. subtilis* 168, *M. tuberculosis* H37Rv, and *P. aeruginosa* PAO1 found in our literature review described above (Table 1).

We analyzed these datasets to quantify the overlap among them. We found that the first regulatory dataset reported for *M. tuberculosis* H37Rv (18) is completely included in the second regulatory dataset reported (19), so we discarded the former. Interestingly, we discovered that the remaining independent regulatory datasets share a poor amount of genes and interactions (Figure 4). The lack of redundancy between them offers an opportunity to obtain more complete RN models through meta-curation. Briefly, we translated the regulatory datasets into locus tags using the constrained mapping strategy described above. We then merged the datasets avoiding duplicating genes and interactions. Regulatory effects and evidences supporting interactions were inherited, if available, from the original datasets. Finally, the meta-curated datasets were processed by the pipeline for data integration as a normal dataset for their inclusion in Abasy Atlas.

These novel meta-curated datasets represent the most complete RNs currently reported for *B. subtilis* 168, *M. tuberculosis* H37Rv, and *P. aeruginosa* PAO1, as determined by their genomic coverage: 75%, 77%, and 16%, respectively. Remarkably, these meta-curated regulatory networks, regardless of their coverage, exhibit properties that characterize hierarchical-modular RNs (see their global network properties in Abasy Atlas) (10,11). All this is suggestive of the good quality of these novel meta-curated RN models that is inherently dependent on the quality of the source datasets.

Datasets obtained from RegTransBase were filtered to reconstruct only the best RN models and to eliminate redundancy from Abasy Atlas as follows. After modeling and analyzing the 477 datasets obtained from RegTransBase, we retained 115 datasets and discarded the remaining 362 for being very small networks (less than 110 genes) mostly composed by disconnected, small (around five genes on average), tree-like structures (regulons). During the construction of Abasy Atlas, UniProt team identified a high level of proteome redundancy in unreviewed UniProtKB (TrEMBL) mainly due to the sequencing and submitting of different strains of the same species (http://www.uniprot.org/help/proteome_redundancy). Since release 2015_04, UniProt discarded redundant entries. To remove redundancy from Abasy Atlas we removed the RNs of organisms with redundant proteomes as identified by UniProt. This step affected only RNs obtained from RegTransBase. After these filtering, we retained 37 RNs from RegTransBase that were integrated in Abasy Atlas.

Genomic coverage for datasets coming from RegTransBase range from 24% (*Staphylococcus aureus* USA300) to 2% (*Pseudomonas aeruginosa* PA7). Remarkably, across strains of the same species, genomic coverage extends through a large range. For example, *E. coli* K-12 MG1655 (strong evidences) has a genomic coverage of 42%, whereas *E. coli* K-12 DH10B has only 7%. This is explained by the fact that regulatory data curation is strain specific and mapped to a specific genome sequence. Coding sequences are highly conserved across strains of the same species, but the same is not true for regulatory sequences that are highly plastic (14). Therefore, regulatory data cannot be extrapolated to other strains even if they belong to the same species.

**Extraction of functional information**

To provide a functional context to RN model genes, we collected the annotated product functions, gene ontology (GO) terms and clusters of orthologous genes (COGs). We extracted product functions from NCBI GenBank, RegTransBase and RegulonDB; GO terms from UniProt-GOA (GO annotation) (28), and COGs from EMBL eggNOG (29) (Figure 5 and Supplementary File 1).

**Computation of predictions**

The procedure to populate Abasy Atlas is mostly automated. We developed a Python script (Figure 5) that is responsible for mapping gene symbols onto canonical gene names, collecting accession IDs for cross-linking and gene symbols for search in the Abasy Atlas, computing systems and systems-level elements by implementing the NDA (Figure 1), computing network global properties, and annotating each system identified by the NDA.

Among the global properties computed for a RN model, we computed the degree distribution, out-degree distribution, and clustering coefficient distribution. Commonly these distributions are fitted to a power-law by ordinary least squares, which is a method that is highly sensitive to outliers in data. Instead, we fitted these distributions by using robust linear regression of log-log-transformed data with Huber's T for M-estimation to overcome the negative effect of outliers.

The pipeline annotated each NDA-predicted module by computing the functional enrichment of the cluster of genes. Functional enrichment analysis was carried out by computing *p*-values for each GO term in the cluster of genes using a hypergeometric distribution probability test. We considered only GO terms in the biological process namespace because they reflect physiological processes. We then corrected *p*-values for multiple tests into *q*-values by controlling the false discovery rate (FDR) at level 0.05. We selected UniProt GO annotation as the functional classifications schema for annotating systems because it is actively and manually curated (28) and provides a shared vocabulary among organisms and databases (30).

**A standard for modelling regulatory networks, interaction effects and evidences, and NDA predictions**

We represented our RN models using JSON (JavaScript Object Notation) open standard format (http://www.json.org/), which is a lightweight, language independent, widely used, data-interchange format supported by more than 50 programming languages. JSON is easy for humans to read and write and for machines to parse and generate. These properties also simplified the serialization of RN models, allowing their processing and visualization in interactive network panels.

**Database schema**

A relational database implemented in MySQL models the metadata and predictions for all RN models, including the RN models in JSON format, and the identity and functional information throughout 14 tables (Supplementary File 2). Three tables contain manually curated information (reg_complexes, sources, and source_urls). Five (organisms, genes, gene_ids, modules, computational_annot) were generated by the pipeline (Figure 5), which is responsible for computing NDA predictions, RN model metadata containing global network properties, and systems computational annotation for each RN model. Six tables store functional data extracted from external sources (gos, goa, cogs, cog_families, cog_classes, and cog).

**Utility and discussion**

**Biological diversity in Abasy Atlas**

Abasy Atlas goes beyond current databases of regulatory information by inventorying systems, global network properties, and systems-level elements for 50 reconstructed and meta-curated RNs (78,649 regulatory interactions) covering 42 bacteria (including pathogenically and biotechnologically relevant organisms) in nine taxa, containing 3,708 regulons and 1,776 systems (Figure 6). Given that nowadays there are no models estimating the total number of regulatory interactions in a given RN, we evaluated the completeness of each RN model in terms of its percentage of genomic coverage. The two most-complete (genomic coverage greater than 75%) RN models in Abasy Atlas are our meta-curations for *M. tuberculosis* H37Rv (77%) and *B. subtilis* 168 (75%). Other RN models having a high genomic coverage are *E. coli* K-12 including regulatory RNAs (73%), *E. coli* K-12 all evidences (72%), *B. subtilis* 168 reconstructed from Arrieta et al. (70%), and *M. tuberculosis* H37Rv reconstructed from Minch et al. (62%). We also reconstructed the most complete RN model for *P. aeruginosa* PAO1 by meta-curation, which has a genomic coverage of 16%.

All the RN models in Abasy Atlas have genes in the four classes of systems-level elements predicted by the NDA, except for four organisms whose RN models (currently) do not have intermodular genes: *Streptococcus pyogenes* serotype M18 (strain MGAS8232), *Streptococcus pneumoniae* (strain Hungary19A-6), *Streptococcus pyogenes* serotype M12 (strain MGAS9429), *E. coli* (strain SMS-3-5 / SECEC). Nevertheless, these are quite incomplete organisms having a genomic coverage less than 9%.

**Overview of the Abasy Atlas interface**

Abasy Atlas has a hierarchical structure enabling users to browse through the available RN models, their systems and systems-level elements. In every section of Abasy Atlas, a top menu bar displays five options and a search interface (Figure 7). The options are **Homepage**, **Browse**, **Downloads**, **About**, and **Contact**. **Homepage** is a shortcut to the homepage. **Browse** enables the user to explore through the atlas. **Download** displays an interface to download data as flatfiles. **About** provides background information on the atlas and the NDA. Finally, **Contact** provides a form for feedback from users. The **Browse** option and the search interface provide the two main ways to start exploring data contained in Abasy Atlas. They will be discussed in the following sections along with graphical browsing of RN models.

Abasy Atlas is cross-linked to various external databases and sites providing biological, genomic, and molecular details. When a user visits an external database or site this opens in a new browser tab, whereas hyperlinks redirecting to other sections in Abasy Atlas open in the same tab.

**Browsing Abasy Atlas**

The **Browse** option is the entry point for browsing the whole atlas. This redirects to a section listing all the RN models contained in the atlas. Each RN model lists genome size, RN genomic coverage, PubMed IDs of the data source, number and percentage of global regulators, modular genes, intermodular genes, and basal machinery genes, number of systems (modules) and a link to a separated section listing the global properties of the RN model. From here, the user can list all the genes in the RN model, retrieve a list of all the genes in a given class, or list all the identified systems.

In the list of genes, genes are listed along with their product description and the predicted class of system-level element (the latter only applicable for all and modular genes). Each gene name is canonical (see Construction and content) and links to a gene details section showing identity (locus tag, UniProt ID, NCBI GeneID, and synonyms) and functional (product function, GO terms, and COGs) information on the gene along with their in-degree, out-degree and clustering coefficient, and an interactive network panel displaying the gene and their graph neighborhood (see Interactive network panel). Here the canonical gene name cross-links to specialized databases containing genomic and molecular details.

By following the modules link in the RN models listing, the users get a list of all the systems identified in that RN model (Figure 7). Data displayed in that list comprise the number of genes belonging to that system, the enriched GO terms and their $q$-value, and a module ID linking to another section displaying the system in an interactive network panel and listing the genes belonging to the system along with their product descriptions.

**Search by gene**

An interface to search Abasy Atlas is also available in the top menu bar of every section (Supplementary File 3). Users may search a gene in all the RN models (by default) or in a subset by selecting the proper option in a searchable dropdown list allowing multiple selections. Search using this interface is case-insensitive and the user may employ wildcards in the search string. Supported wildcards are '?' to match any single character and '*' to match any arbitrary number of characters including zero.

The search engine looks for the query in the set of canonical gene names, synonyms and locus tags returning a composite set of results grouped for each RN model and ordered by canonical gene name (Supplementary File 3). The results section lists all the matches found providing canonical gene names, NCBI GeneIDs, locus tags, UniProt IDs, synonyms, and classes of systems-level elements. Gene names link to the details section described above. If the class of systems-level elements is a system, this links to the corresponding entry in the list of all the systems identified in that RN model.

**Interactive network panel**

Abasy Atlas displays gene graph neighborhoods (i.e., the set of all the genes/complexes regulating or being regulated by a given gene, including the latter) and systems in interactive network panels that share a common interface (Figure 8). This interface provides a button to download an image in PNG format with transparency containing the network displayed, which is free for use in papers and other academic/non-commercial uses as long as proper citation is provided (see Availability and requirements). It is also possible to remove global regulators, disconnected nodes, and weakly supported interactions temporally. A dropdown list enables the user to apply a layout to the network, and a button reapplies the selected layout. A checkbox controls if layout transitions are animated or not, this is a useful control when the number of nodes could slow the web browser. The full quality checkbox controls whether full quality is used during network manipulation and animated layout transitions. The animation is off by default, and we suggest this for improved responsiveness. The configuration of all the checkboxes and selected layout is preserved across a session in the same web browser tab. When the user closes the tab, the configurations are restored to default.

We fine-tuned the parameters for rendering the network in order to optimize responsiveness and user experience. Besides, we developed an adaptive layer running on top of the algorithms rendering the network. Before a network is rendered, this adaptive layer assesses the performance of the user's computer and, if necessary, overrides user's selections for a smoother experience. If the adaptive layer found that the user's computer will exhibit degraded performance for a particular network, then layout animation is disabled and the network is always initially displayed using the grid layout despite previous user selections. The user can always re-enable layout animation or change the layout upon accepting a warning regarding the risk of degraded performance.

In the upper-right corner of the network panel, a widget similar to that used by Google Maps is available. This widget enables panning and zooming the network. The user may also use the mouse wheel for zooming the network. Inside the interactive panel, the network is displayed by using a color code for nodes and interactions. We encoded interaction effects in line colors and arrowhead shapes as follows: red and T shape for repressions, green and arrow shape for activations, orange and diamond shape for dual, and grey and arrow shape for unknown effect. Evidences supporting interactions, if available, are encoded as different line styles: solid for strong evidences and dashed for weak evidences. Genes belonging to the same system or class of systems-level element have the same color. In the gene details section, the central gene is highlighted in yellow. If the central gene product is a subunit shared by several complexes, all these complexes are also highlighted. We represented heteromeric regulatory complexes as pentagons, whereas genes are circular nodes.

All the nodes can be draged around and repositioned, and clicking on an interaction switches its state between normal and highlighted. If the mouse hovers over a node or interaction, a tooltip displays providing information such as local (neighborhood dependent) out- and in-degrees, subunits if the node is a regulatory complex, and links to other genes and systems. The user also may explore the RN model by graphical browsing. If the user clicks in a node, he is redirected to the details section of the corresponding gene.

**Use case 1: Global regulators of *C. glutamicum* ATCC 13032**

The NDA defines an equilibrium point (κ-value) between two apparently contradictory behaviors occurring in hierarchical-modular networks: hubness and modularity. In these networks, modularity is inversely proportional to hubness (Figure 1). The κ-value is defined as the connectivity value for which the variation of the clustering coefficient (modularity) equals the variation of out-connectivity (hubness) but with the opposite sign (dC/dk$_{out}$ = −1) (9,10). This is a network-dependent parameter that allows the identification of global regulators as those nodes with connectivity greater than κ. We identified 19 global regulators in the *C. glutamicum* RN model (κ-value = 11): *sigA*, *glxR*, *sigH*, *dtxR*, *ramA*, *ramB*, *lexA*, *mcbR*, *cg2115* (*sugR*), *amtR*, *mtrA*, *cgtR11* (*hrrA*), *ripA*, *cgtR3* (*phoR*), *cg0156* (*cysR*), *sigM*, *cg1324* (*rosR*), *cg0196* (*pckR*), and *sigB*. These global regulators comprise 4 out of the 7 sigma factors encoded in the genome of *C. glutamicum* (31). All of these regulators have been jointly described as global regulators (32) except for *mtrA*, *cgtR11* (*hrrA*), *ripA*, *cg0156* (*cysR*), *sigM*, *cg1324* (*rosR*), and *cg0196* (*pckR*). Nevertheless, *cgtR11* (*hrrA*) (33) has also been individually reported as global regulator. In our set of global regulators we only missed *arnR* (*cg1340*) that has been reported as a global regulator (32), but it is a modular gene belonging to a system annotated as "nitrate metabolic process" according to the NDA prediction.

**Use case 2: Guilt by association, the case of *B. subtilis* 168 *hemD*, *yppF*, *yptA* and *yycS* involved in the biosynthesis of antibiotics and protoporphyrinogen IX**

Genomes are functionally annotated by using homology analysis. Annotation is conducted by sequence similarity as assessed by BLAST-based tools, and it is complemented with structural features and comparative genomic context analysis. Nowadays, genome annotation has reached a mature development but yet there are no functional predictions for hundreds of genes (e.g. those annotated as hypothetical predicted/conserved protein) even in the best characterized genomes because no homologs have been characterized (34). Genomic context analysis is based on the idea of exploiting all types of functional associations between genes in the same or in different genomes that may indicate a common function justifying a verdict of guilt by association.

Abasy Atlas goes beyond genomic functional associations providing systems-level functional associations allowing identification of the function of genes whose annotation by homology analysis is not possible. In Abasy Atlas, 682 genes annotated as hypothetical or with missing annotation belong to functionally annotated systems. This provides a way to assign functional annotations to these genes by using a guilt-by-association strategy: if gene G belongs to a system with function X, then G is also involved in X.

For example, *hemD*, *yppF*, *yptA* and *yycS* are genes belonging to the system 1.19 of our meta-curated model for *B. subtilis* 168 (Figure 8). This system is annotated with the GO terms "Protoporphyrinogen IX biosynthetic process" (GO:0006782), "Metal ion transport" (GO:0030001), and "Antibiotic biosynthetic process" (GO:0017000). This provides evidence from a systems-level context to propose that these genes are also involved in these functions.

**Data available for download**

To facilitate the access to and analysis of the RN models and the systems predicted, we provide various datasets including the compendium of genes names, synonyms and accessions IDs as flat files for download. The datasets available are:

1. RN models in JSON data-interchange format, including NDA predictions and, if available, effect and evidences supporting regulatory interactions.
2. Gene information as follows: canonical gene name, locus tag, NCBI GeneID, UniProt ID, synonyms, product function, and class of systems-level element predicted by the NDA.
3. Module annotation comprising module ID, gene ontology term, and *q*-value supporting the annotation.

**Future development plans**

The long-term goal of Abasy Atlas is to provide an integrated platform to perform comparative large-scale studies of bacterial regulatory systems and the functional architectures governing them from a large-scale systems biology perspective. To accomplish this goal, we envision several improvements and extensions:

- Improvements to the search interface and engine to provide support for using regular expressions and searching by function.
- Expansion to include the regulators and systems controlling the intermodular genes, and the regulators of the basal machinery genes.
- Expanding the global properties section to include new interesting properties.
- Introduction of an interactive network panel to browse a RN model as a functional architecture displaying the hierarchy governing the systems-level elements.
- Providing an online tool for applying the NDA to a RN uploaded by the user to obtain its system-level elements and functionally annotated systems.
- Classification of the force of evidences supporting regulatory interactions in other organisms.
- Updating and expanding our RN models by meta-curation of new datasets and inclusion of predicted regulatory datasets as those from RegPrecise (35), PRODORIC (36), and CoryneRegNet (4).
- Enabling the user to explore gene expression in the interactive network panel at the level of gene neighborhood, system, and functional architecture by mapping gene expression high-throughput data uploaded by the user onto the RN model.

**Conclusions**

Abasy Atlas takes the first step towards a global understanding of the RNs organization by providing a comprehensive inventory of annotated functional systems, global network properties, and systems-level elements across a large range of bacteria, including organisms of medical and biotechnological relevance. In addition, meta-curation of regulatory datasets provides the most complete and reliable set of regulatory interactions currently available, which can be projected into subsets by considering the force or weight of evidence supporting them or the systems that they belong to.

The NDA provides formal and quantitative criteria to identify the systems and systems-level elements shaping the functional architecture of RNs. The decomposition of diverse RNs into their systems and systems-level elements beyond the regulon allows unraveling the complexity of these networks and provides new insights into the organizational principles governing them. The prevalence of intermodular genes (a novel system-level element first identified by the NDA) across a large range of prokaryotes support the diamond-shaped, three-layer, hierarchy unraveled by the NDA as a universal organizational principle among bacteria.

Abasy Atlas provides a standardized framework for biological networks annotation and meta-curation. For example, the gene-name disambiguation algorithm may be applied to other projects to reduce the influence of this confounding factor. Abasy Atlas also provides data enabling future large-scale comparative systems biology studies to understand the common principles and particular lifestyle adaptations of regulatory systems across bacteria. All this brings together a large body of data that will surely inspire studies to generate hypothesis regarding the principles underpinning the evolution and organization of systems and the functional architectures controlling them.

**Availability and requirements**

Abasy Atlas is available for web access at http://abasy.ccg.unam.mx. If you use any material from Abasy Atlas please cite properly. Use of Abasy Atlas and each downloaded network PNG image by Ibarra-Arellano et al. is licensed under a Creative Commons Attribution 4.0 International License. Permissions beyond the scope of this license may be available at jfreyre@ccg.unam.mx. **Disclaimer**: Please note that original data contained in Abasy Atlas may be subject to rights claimed by third parties. It is the responsibility of users of Abasy Atlas to ensure that their exploitation of the data does not infringe any of the rights of such third parties.

**List of abbreviations**

RN, regulatory network; NDA, natural decomposition approach; JSON, JavaScript Object Notation; GO, gene ontology; COG, cluster of orthologous genes

**Competing interest**

The authors declare that they have not competing interest.

**Authors' contributions**

MAI-A and AIC-G developed software for the automated data extraction, parsers and converters, and carried out the meta-curation of RNs. MAI-A and JAF-G developed software for NDA predictions and automated functional annotation. AIC-G developed software for the disambiguation and mapping of gene symbols into canonical gene names. AT provided *C. glutamicum* regulatory interaction data. LGT-Q and AT provided biological expertise. JAF-G conceived and designed Abasy Atlas, coordinated its development, developed the interactive network panel and the final version of the web interface and database, and wrote the manuscript. All authors read and approved the final manuscript.

**Authors' information**

MAI-A holds a BSc in Genomic Sciences and is research intern at University of Florida. AIC-G is senior student at the Undergraduate Program in Genomic Sciences, UNAM. LGT-Q holds a PhD in Biochemistry and a BSc in Pharmaco-Biological Chemistry and is professor of Environmental Microbiology and head of the group of Contamination and Sustainability at the Polytechnic University of the State of Morelos. AT holds a PhD in Genetics and is the co-coordinator of the German Network for Bioinformatics Infrastructure – de.NBI. JAF-G holds a PhD in Biochemistry, an MSc in Computer Science, and a BSc in Computer Systems Engineering and is associate professor of systems biology and head of the group of Regulatory Systems Biology at the Center for Genomic Sciences, UNAM.

**Funding**

This work was supported by the Programa de Apoyo a Proyectos de Investigación e Innovación Tecnológica (PAPIIT-UNAM) [IA200614 and IA200616 to JAF-G].

**Acknowledgements**

We thank Josue Jiménez for drafting the first version of the web interface and database, and José Enrique León-Burguete and Aldo Carmona for technical support and manual curation. We thank Ricardo Ramírez-Flores and Patricia Romero-Nájera for technical support and fruitful discussions. We also thank two anonymous reviewers for helpful suggestions.

**Figures**

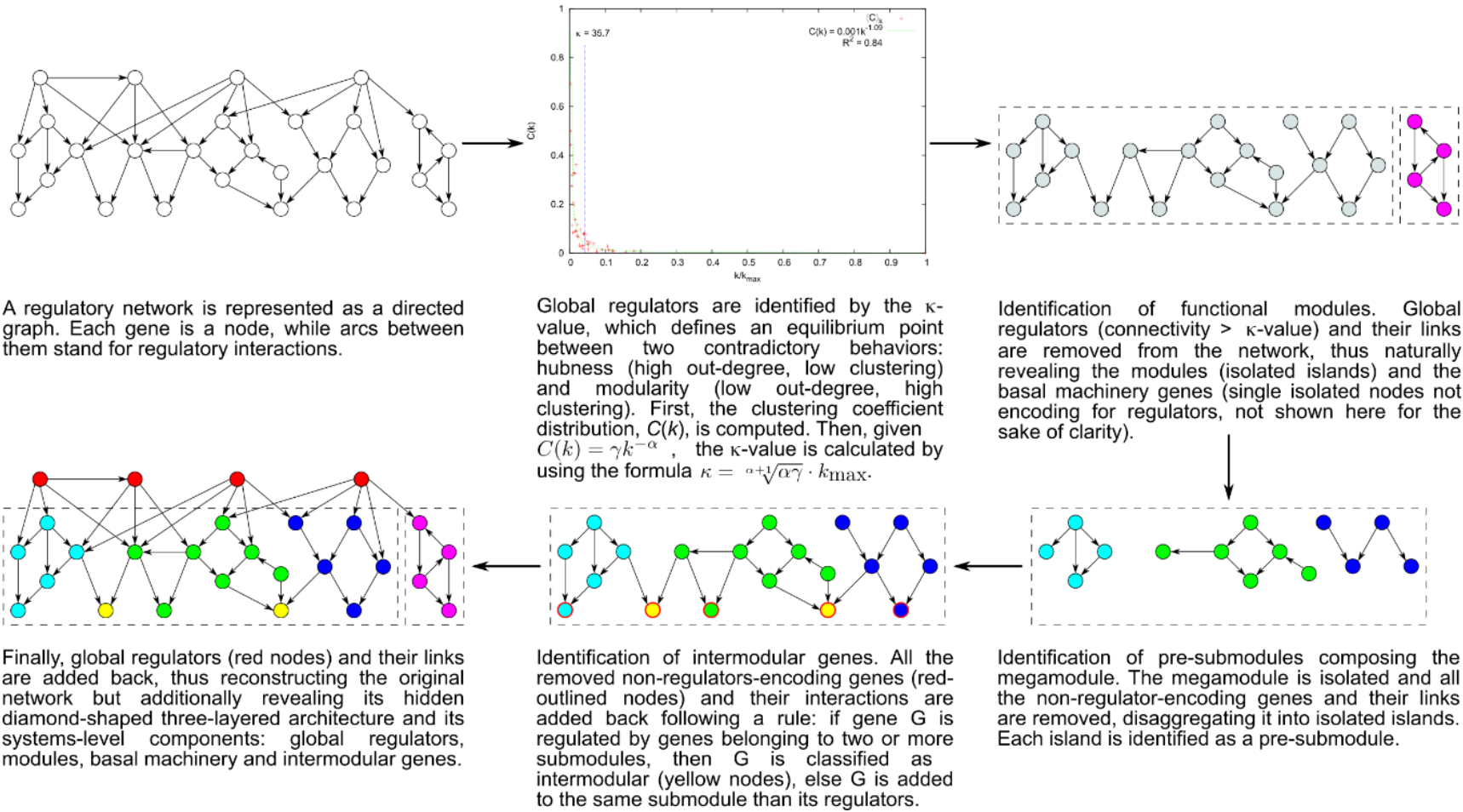

Figure 1. The natural decomposition approach.

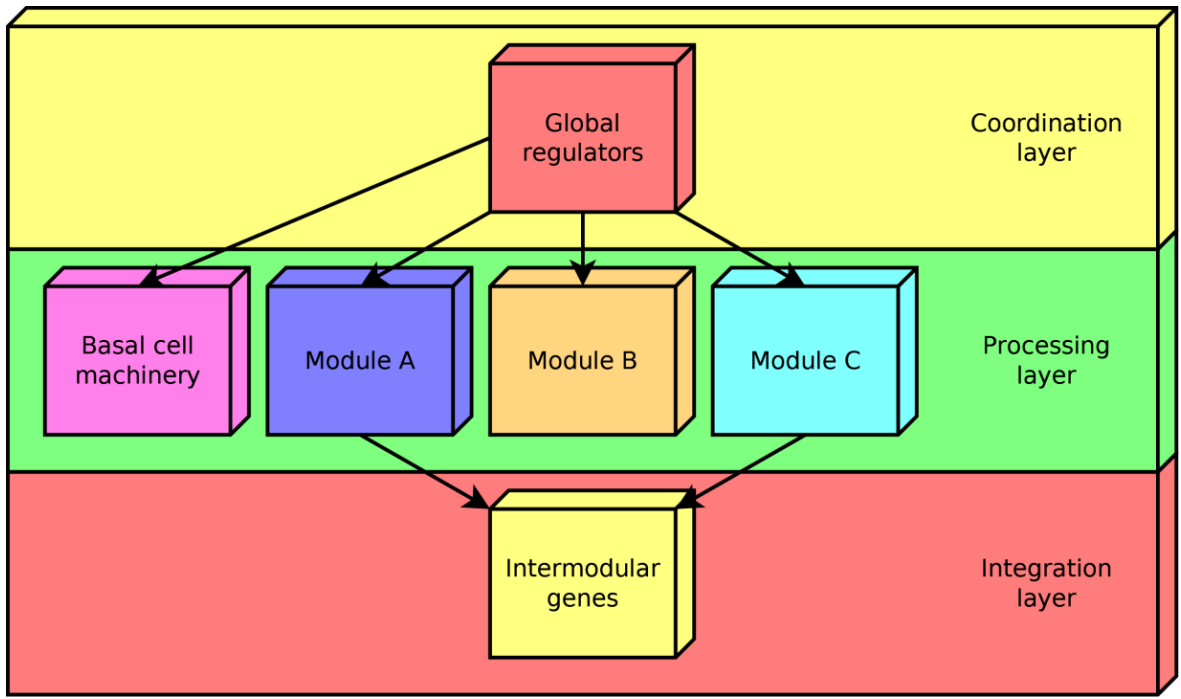

Figure 2. Common functional architecture identified by the NDA in bacteria. The functional architecture unveiled by the NDA is a diamond-shaped, three-layer, hierarchy, exhibiting some feedback between processing and coordination layers, which is shaped by four classes of systems-level elements: global transcription factors, locally autonomous modules, basal machinery and intermodular genes.

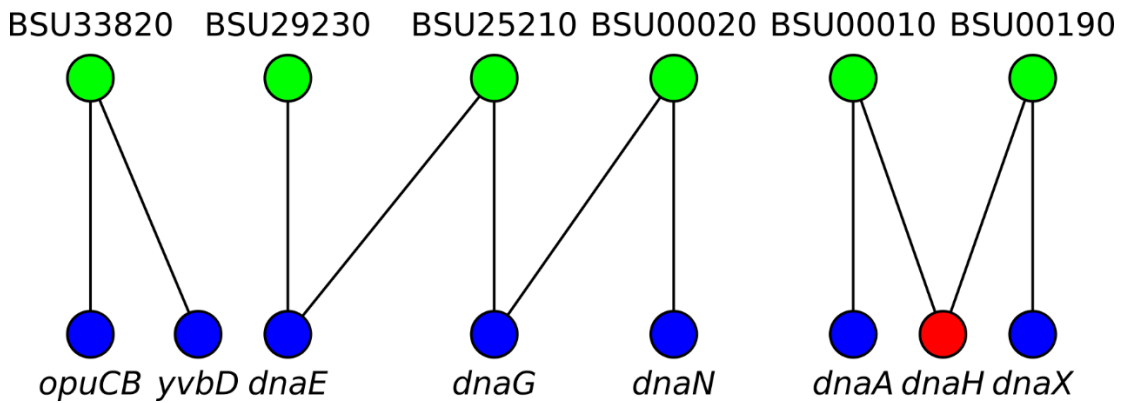

Figure 3. Gene names disambiguation algorithm. A bipartite graph models the equivalence relations between locus tags (upper nodes) and gene names (lower nodes). Canonical gene names and locus tags relate via vertical links, whereas diagonal links connect synonyms and locus tags. Mapping from gene names into locus tags can follow any vertical link but diagonal links are taboo (red node) except if node degree is exactly one (e.g. *yvbD*). In this example, *opuCB*, *dnaE*, *dnaG*, *dnaN*, *dnaA* and *dnaX* are solved via canonical gene names; *yvbD* is found to be a synonym for *opuCB* and both are the same gene; whereas *dnaH* cannot be solved because it exhibits degeneracy (node degree greater than one, and no vertical link denoting it as a canonical gene name).

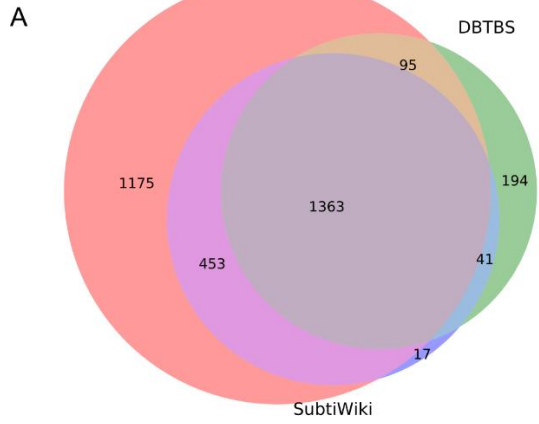
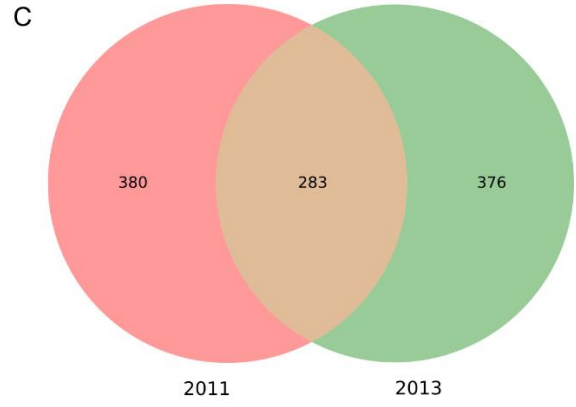
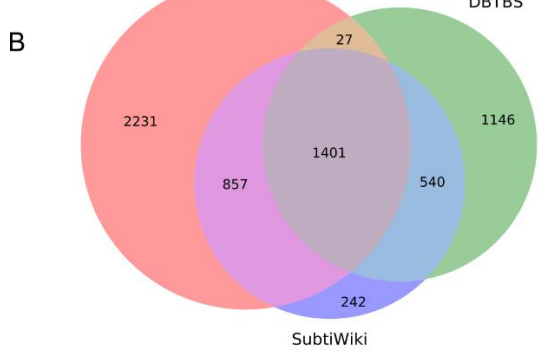
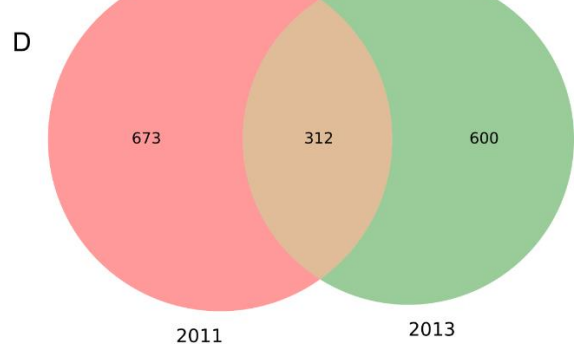
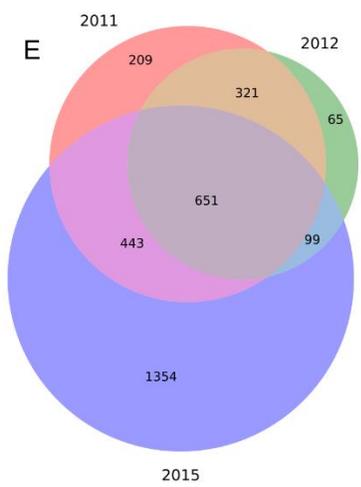
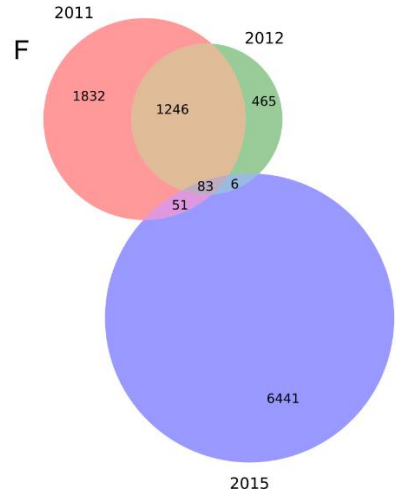

Figure 4. Meta-curation of regulatory datasets. Poor overlap between different datasets for (A) nodes and (B) interactions in *B. subtilis* 168, (C) nodes and (D) interactions in *P. aeruginosa* PAO1, and (E) nodes and (F) interactions in *M. tuberculosis* H37Rv.

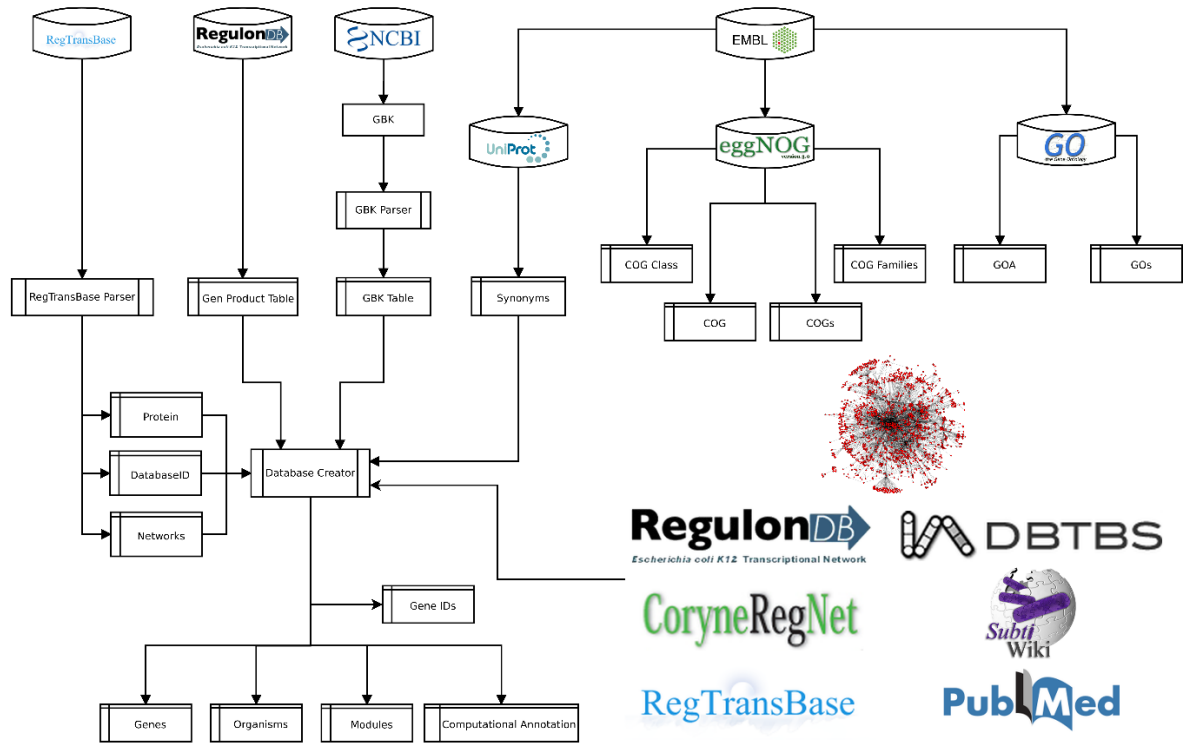

Figure 5. Pipeline for data integration.

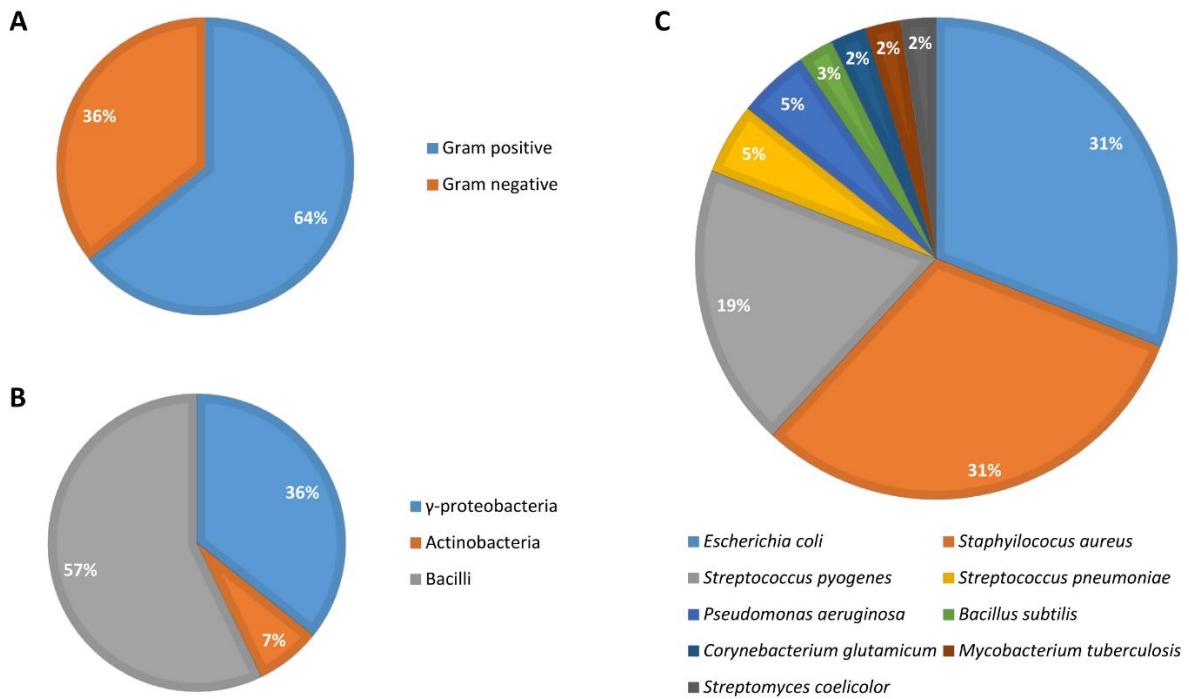

Figure 6. Biological diversity in Abasy Atlas.

Figure 7. Listing of the systems identified in our meta-curation of *M. tuberculosis* H37Rv.

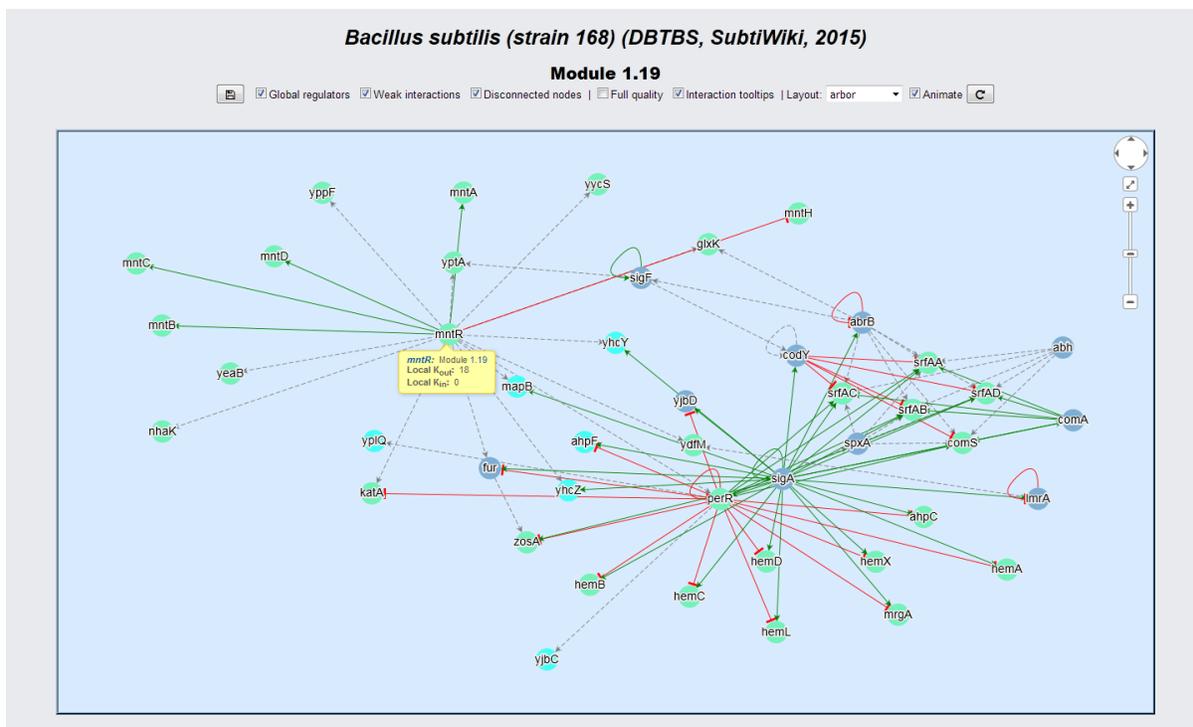

Figure 8. Interactive network panel for module 1.19 of our meta-curation of *B. subtilis* 168. For a description of the meaning of different line-styles and colors, please refer to the section "Interactive network panel" in the main text.

**Tables**

Table 1. Regulatory datasets used for meta-curation.

| Regulatory dataset | Organism | Genomic coverage |
|---|---|---|
| Balazsi et al. (2008) | *M. tuberculosis* H37Rv | 22.9% |
| Sanz et al. (2011) | *M. tuberculosis* H37Rv | 39.7% |
| Rohde et al. (2012) | *M. tuberculosis* H37Rv | 27.8% |
| Minch et al. (2015) | *M. tuberculosis* H37Rv | 62.1% |
| Galan-Vasquez et al. (2011) | *P. aeruginosa* PAO1 | 11.6% |
| RegTransBase - Cipriano et al. (2013) | *P. aeruginosa* PAO1 | 11.6% |
| DBTBS - Sierro et al. (2008) | *B. subtilis* 168 | 38.2% |
| SubtiWiki - Michna et al. (2016) | *B. subtilis* 168 | 42.4% |
| Arrieta-Ortiz et al. (2015) | *B. subtilis* 168 | 69.8% |

**Additional files**

Supplementary File 1. Sources for data.

Supplementary File 2. Database schema.

Supplementary File 3. Querying Abasy Atlas.